\documentclass[12pt]{article}
\pdfoutput=1
\usepackage{amsmath} 
\usepackage{longtable}
\usepackage{graphicx,epsfig}    
\usepackage{color}
\usepackage{cite}
\usepackage{amssymb}
\usepackage[utf8]{inputenc}
\usepackage{appendix}
\usepackage{array}
\definecolor{darkblue}{RGB}{0, 56, 128}
\usepackage[hyperfootnotes=true,bookmarks=false,
hypertexnames=true,pagebackref=false,linktocpage=false,hidelinks]{hyperref}
\hypersetup{
   colorlinks=true,%
   linkcolor=darkblue,%
   citecolor=darkblue,%
   filecolor=green,%
   menucolor=cyan,%
   urlcolor=darkblue}

\usepackage{footnotebackref}

\newcommand{\as}{\alpha_s}

\newcommand{\astwopi}{\left(\frac{\alpha_s}{2 \pi}\right)}

\newcommand{\Duno}{\mathcal{D}_1 (x)}

\newcommand{\dif}[1]{\;{\rm d}{#1}\,}

\def\P{\mathcal{P}}
\def\B{\mathcal{B}}
\def\T{\mathcal{T}}
\def\R{\mathcal{R}}
\def\V{\mathcal{V}}
\def\M{\mathcal{M}}
\def\G{\mathcal{G}}
\def\Q{\mathcal{Q}}
\renewcommand{\O}{\mathcal{O}}

\begin{document}
\setlength{\parskip}{0.15cm}
\setlength{\baselineskip}{0.52cm}
\begin{titlepage}

\begin{flushright}
ICAS 046/19\\
ZU-TH 51/19\\
MPP-2019-245
\end{flushright}

\renewcommand{\thefootnote}{\fnsymbol{footnote}}

\thispagestyle{empty}
\noindent

\vspace{0.5cm}

\begin{center}
{\bf \Large 
Triple Higgs production at hadron  \\[1ex] colliders at NNLO in QCD\\}
  \vspace{1.25cm}
{\large
Daniel de Florian$\,^{(a)}$\footnote{deflo@unsam.edu.ar},
Ignacio Fabre$\,^{(a,b)}$\footnote{fabre@physik.uzh.ch} and 
Javier Mazzitelli$\,^{(c)}$\footnote{jmazzi@mpp.mpg.de} \\
}
 \vspace{1.25cm}
 {
    $^{(a)}$ International Center for Advanced Studies (ICAS) and ICIFI, ECyT-UNSAM,  \\
25 de Mayo y Francia, (1650) Buenos Aires, Argentina\\[0.3cm]
$^{(b)}$ Physik-Institut, Universit\"at Z\"urich, \\ 
Winterthurerstrasse 190, CH-8057 Z\"urich, Switzerland\\[0.3cm]
$^{(c)}$ Max-Planck-Institut f\"ur Physik, F\"ohringer Ring 6, 80805 M\"unchen, Germany
\\
 }
  \vspace{1.5cm}
  \large {\bf Abstract}
  \vspace{-0.2cm}
\end{center}

In this work we analyse the triple Higgs boson production cross section at hadron colliders via gluon fusion, and present for the first time the full set of QCD NNLO corrections in the heavy top limit. In order to account for finite top mass effects we perform two different reweighting procedures,
and study the dependence of the result on the choice of the approximation.
Combining the most accurate predictions available to date, we present the following result for the total NNLO cross section for triple Higgs boson production at a 100~TeV collider:  $\sigma_\text{NNLO}=5.56^{+5\%}_{-6\%} \pm 20\%$~fb, where the first uncertainty is an estimate for higher order effects from scale variations, while the last one is an estimate for the missing finite top mass effects. 
\hfill

\end{titlepage}
\setcounter{footnote}{1}
\renewcommand{\thefootnote}{\fnsymbol{footnote}}
%
\section{Introduction}
\label{sec:intro}
The discovery of the Higgs boson in 2012~\cite{Aad:2012tfa,Chatrchyan:2012xdj} marked a milestone for the particle physics community. Since then, great efforts have been made to measure its properties in order to determine if it corresponds to the Standard Model (SM) Higgs boson or if it opens a window to new physics beyond it.
One of the crucial aspects of this quest is the determination of its self-couplings, which are directly related to the scalar potential that drives electroweak symmetry breaking, and can be modified in the presence of new-physics effects.

In the SM, the triple and quartic self couplings of the Higgs, $\lambda_{3}$ and $\lambda_4$, are uniquely fixed by its mass $m_H$ through the enforcement of gauge symmetries and renormalisability of the theory, namely $\lambda_{3} = \lambda_{4} =\lambda_{\text{SM}} = m_H^2 /(2 v^2)$, where $v \approx 246$~GeV is the Higgs vacuum expectation value. These couplings are directly accessible through double (for $\lambda_{3}$) and triple (for $\lambda_{4}$) Higgs boson production, although they might also be extracted indirectly from single Higgs production~\cite{McCullough:2013rea,Gorbahn:2016uoy,Degrassi:2016wml,Bizon:2016wgr,DiVita:2017eyz,Maltoni:2017ims} and precision electroweak observables~\cite{Degrassi:2017ucl,Kribs:2017znd}. 
All of these measurements, however, are extremely challenging, and in particular already the determination of the triple self-coupling will prove very difficult at the high-luminosity phase of the LHC (see Ref.~\cite{DiMicco:2019ngk} for a review).
Unfortunately, the much smaller triple-Higgs production rate makes the determination of $\lambda_4$ prohibitive at the LHC, and even challenging at future colliders~\cite{Papaefstathiou:2015paa,Papaefstathiou:2019ofh}.
New physics effects, however, might change the prospects of measuring triple-Higgs production by inducing large enhancements of the cross section, and in this respect it is therefore desirable to provide precise predictions for the SM expectation. 

As it occurs for single and double Higgs, the dominant triple Higgs production mechanism at hadron colliders is gluon fusion, mediated by a heavy-quark (mostly top quark) loop. This process, being gluon induced, is expected to present large QCD corrections. However, due to the massive loop appearing in the amplitude at Born level and the large number of external particles, the calculation of its higher order corrections is extremely challenging. As a consequence, the exact cross-section for triple Higgs production is only known at leading order (LO) in QCD~\cite{Plehn:2005nk,Binoth:2006ym}, while next-to-leading order (NLO) corrections are currently unknown.
In order to provide precise predictions for triple Higgs production, approximate NLO corrections were calculated in Ref.~\cite{Maltoni:2014eza} within the heavy top limit (HTL) for the virtual amplitudes, while keeping the exact dependence on the top mass $m_t$ for the real emission diagrams (this approximation is denoted FT$_\text{approx}$). In Ref.~\cite{deFlorian:2016sit} the NNLO virtual amplitudes where computed in the HTL, and the so-called soft-virtual approximation of the HTL NNLO corrections (NNLO$_\text{SV}$) was obtained~\cite{deFlorian:2012za}. Phenomenological results were presented by reweighting the NNLO$_\text{SV}$ cross-section by the exact LO result (i.e. with full $m_t$ dependence)~\cite{deFlorian:2016sit}, in what is known as the \emph{Born-improved} (Bi) cross-section.
The main goal of the present work is to extend the results of Ref.~\cite{deFlorian:2016sit} beyond the soft-virtual approximation, computing the complete set of NNLO QCD corrections for triple Higgs production within the HTL.
In addition to that, we include partial finite-$m_t$ effects by taking into account the NLO FT$_\text{approx}$ results of Ref.~\cite{Maltoni:2014eza}, thus providing a final estimate of the cross section that combines the most advanced results available to date.

The paper is organised as follows. In Section~\ref{sec:LO} we examine the structure of the LO triple Higgs production cross-section and its dependence on $\lambda_{3}$ and $\lambda_{4}$. Then, in Section~\ref{sec:correction} we complete the calculation of the NNLO corrections in the HTL by adding the real emission corrections to the results presented in Ref.~\cite{deFlorian:2016sit}. 
Then, in Section~\ref{sec:pheno} we present the phenomenological results for triple Higgs production at the LHC and future hadron colliders.
We estimate the dependence on the reweighting method by comparing the \emph{Born-improved} result with a modified prescription (introduced in Ref.~\cite{deFlorian:2017qfk}), that we call \emph{dynamically Born-improved}.
Finally, we combine our result with the one presented in Ref.~\cite{Maltoni:2014eza} to report our best prediction, and in Section~\ref{sec:conc} we present our conclusions.

\section{The Amplitude at LO}
\label{sec:LO}
In this section we will examine the structure of the LO amplitude and cross-section. The Born amplitudes needed for the numerical calculation were obtained using {\sc Recola2}\cite{Denner:2017wsf}. For the parton distribution functions we adopted the MMHT2014\cite{Harland-Lang:2014zoa} set interfaced via {\sc LHAPDF}\cite{Buckley:2014ana}, while the {\sc CUBA}\cite{Hahn:2004fe} library was used to perform the numerical integration. The  values implemented for the physical input parameters are $G_F = 1.16656 \times 10^{-5}$~GeV$^{-2}$ for the Fermi constant, $m_H = 125$~GeV, $m_t=173.2$~GeV and $\Gamma_H = \Gamma_t = 0$ for the masses and widths of the Higgs boson and the top quark, respectively, and $\alpha_S(m_Z) = 0.135$ for the strong coupling constant at LO, as provided by the MMHT2014 set. Throughout this work, the on-shell top quark mass scheme is used. All the plots in this section correspond to a collider centre of mass (CM) energy of 100~TeV, although we explicitly checked that all our conclusions also hold at 14 and 27~TeV. 

\begin{figure}
    \centering
    \includegraphics[width=.9\columnwidth]{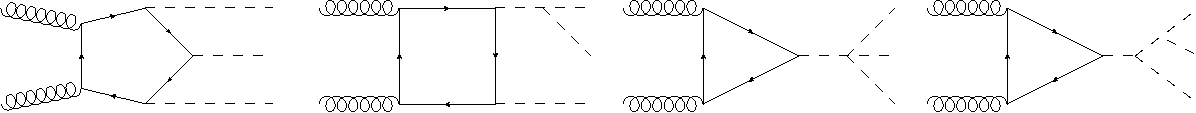}
    \caption{Diagrams that contribute to the LO triple Higgs production in the gluon fusion channel (modulo permutations of the final state bosons). We identify the pentagon $\P$ (left), box $\B$ (second from left) and triangles $\T_4$ and $\T_3$ (right) diagrams, respectively.}
    \label{fig:Born_Diagrams}
\end{figure}

For triple Higgs production, the relevant diagrams (modulo permutations of the final state particules) are shown in Figure~\ref{fig:Born_Diagrams}. We can split them in four different categories: pentagons ($\P$), boxes ($\B$) and two triangle contributions ($\T_1$ and $\T_2$), each one of these with a specific dependence on the parameters $\kappa_3 = \lambda_{3}/\lambda_{\text{SM}}$ and $\kappa_4 = \lambda_{4}/\lambda_{\text{SM}}$ that parametrise departures of the self couplings from the SM expectations,
\begin{align}
    \mathcal{M} &= \P + \kappa_3\,\B + \kappa_3^2\,\T_3 + \kappa_4\,\T_4. \label{eq:LO-contributions}
\end{align}
As in the case of double Higgs production \cite{Glover:1987nx}, there are only two independent helicity configurations of the initial gluons, that we call
\begin{align*}
    \mathcal{M}_{++} &= \mathcal{M}_{--}, \qquad \text{``Spin 0"},\\
    \mathcal{M}_{+-} &= \mathcal{M}_{-+}, \qquad \text{``Spin 2"},
\end{align*}
according to the value of total spin along the collision axis.

\begin{figure}
    \centering
    \includegraphics[width = \columnwidth]{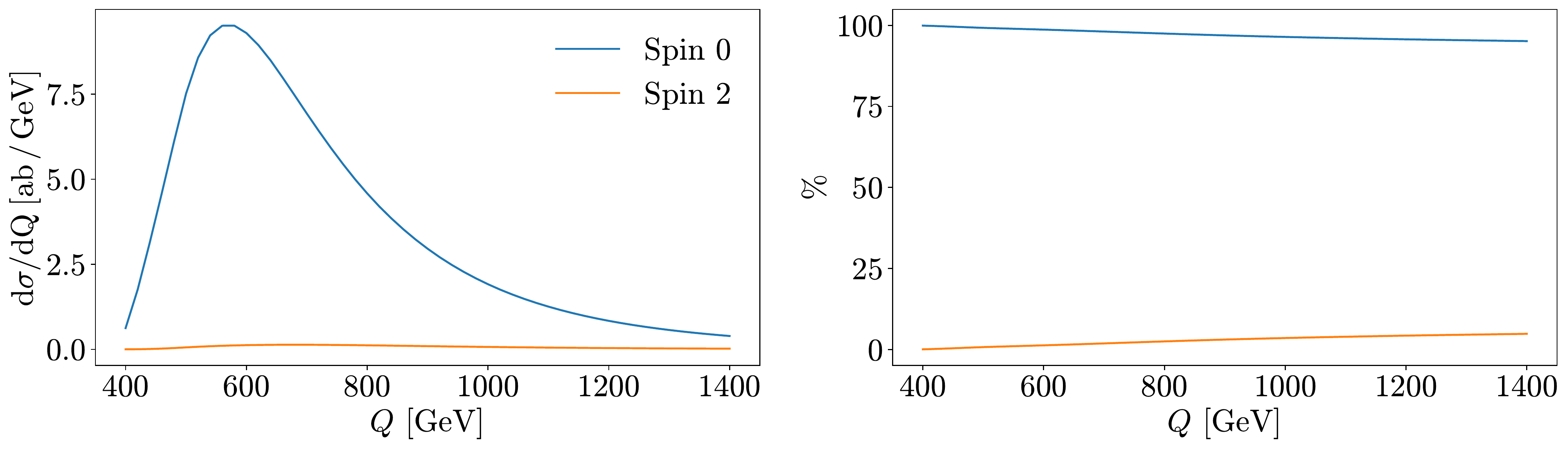}
    \caption{\emph{Spin 0} ($\sim|\M_{++}|^2$) and \emph{Spin 2} ($\sim|\mathcal{M}_{+-}|^2$) contributions to the invariant mass distribution of the triple Higgs system for collider CM energies of 100~TeV. The left plot shows the absolute contributions, while the right one shows the percentage.}
    \label{fig:LO-spin}
\end{figure}

The \emph{Spin 2} configuration vanishes in the limit $m_t\to\infty$, while the \emph{Spin 0} configuration remains. We observe in Figure~\ref{fig:LO-spin} that the contribution from the \emph{Spin 2} piece is rather small (below 5\% of the total cross-section). Also, we notice that triangle contributions $\T$ only contribute to the \emph{Spin 0} helicity configuration, and therefore the \emph{Spin 2} configuration is not sensitive to the quartic Higgs self coupling $\kappa_4$. 

\begin{figure}
    \centering
    \includegraphics[width = \columnwidth]{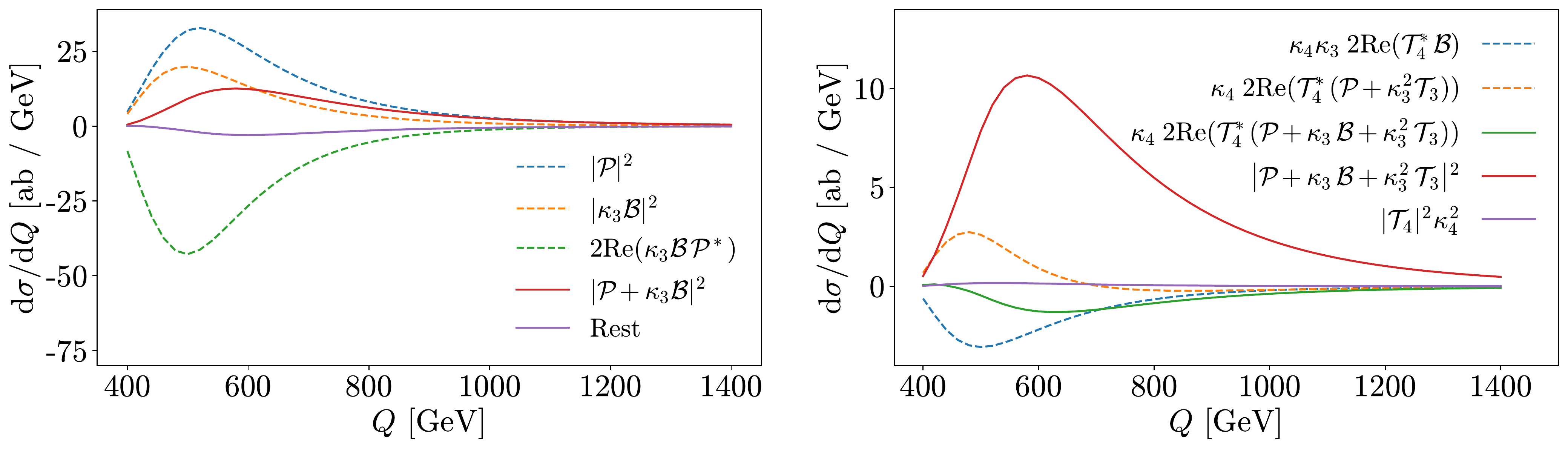} 
    \caption{Different contributions to the \emph{Spin 0} component of the invariant mass distribution. Left:  In dashed lines we show the contributions from the pentagon and box diagrams, as well as their destructive interference, resulting in the red solid line. Right: The contributions separated by their $\kappa_4$ dependence. In dashed lines we show a similar destructive interference pattern, due to the sign difference between the box and the pentagon (and triangle) form factors, resulting in the green solid line.}
    \label{fig:LO-contributions}
\end{figure}

It is interesting to observe the share of the cross-section from each topological contribution and their corresponding interferences. In Figure~\ref{fig:LO-contributions} we show different contributions of the \emph{Spin 0} component to the invariant mass distribution of the triple Higgs system. In the left panel we plot the interference structure between the $\P$ and the $\B$ diagrams, similar to the one usually presented for double Higgs production between the box and triangle contributions. This pattern can be  better understood if we parametrise the amplitudes in terms of quark loop form factors (factoring out the Higgs boson couplings and propagators) and look at their behaviour in the HTL:
\begin{align}
    \T_4 &=:  \frac{\as}{2\pi}\frac{Q^2}{3v^3}\,\frac{3 m_H^2}{Q^2-m_H^2} \, F_\T(Q^2)\nonumber \\
    \T_3 &=: \frac{\as}{2\pi}\frac{Q^2}{3v^3}\, \frac{\left(3 m_H^2\right)^2}{Q^2-m_H^2} \, \sum_{(ij)}\frac{1}{s_{ij}-m_H^2} \,F_\T(s_{ij})\nonumber \\
    \B &=: \frac{\as}{2\pi}\frac{Q^2}{3v^3}\, 3 m_H^2 \sum_{(ij)}\frac{1}{s_{ij}-m_H^2}\,F_\B(Q^2,(p_i+p_j-p_1)^2,(p_i+p_j-p_2)^2,s_{ij}, m_H^2) \nonumber\\
    \P &=: \frac{\as}{2\pi}\frac{Q^2}{3v^3} \,F_\P(p_1, p_2, p_3, p_4, p_5) \nonumber \\
    \implies \lim_{m_t\to\infty} F_\T &= \frac{1}{2}\lim_{m_t\to\infty} F_\P = -\lim_{m_t\to\infty} F_\B = 1,\label{eq:LO-FormFactors}
\end{align}
where $p_1$ and $p_2$ is the four-momenta of the initial gluons, and $p_i$ ($i=3,4,5$) the ones of the outgoing Higgs bosons, $Q^2 = (p_1+p_2)^2$ and $s_{ij} = (p_i+p_j)^2$ are the squared invariant masses of the triple Higgs boson system and of the different pairs $(ij)$, respectively, and $\sum_{(ij)}$ is a sum over the three different pairs $\{(34),(45),(53)\}$.
The dominant contributions come from the pentagon and the box diagrams, as they are less suppressed by Higgs propagators, although the difference in sign between the $F_\P$ with $F_\B$ is responsible for a large negative interference between them. 
This effect is particularly strong at threshold, and therefore this region presents an enhanced sensitivity to any BSM effect that can alter the delicate cancellation between diagrams present in the SM.

In the right panel of Figure~\ref{fig:LO-contributions} we present, in solid lines, the contribution originated from different powers of $\kappa_4$.  The quadratic dependence arises from the $\T_4$ triangle diagram, which is suppressed by a Higgs propagator ( $\sim Q^{-2}$) making it negligible, with a contribution at the peak of $\sim 1.5$\%. The linear term has two contributions, plotted as dashed lines, that cancel each other to a large extent. This cancellation is due to the sign difference of the form factors in Eq.~\eqref{eq:LO-FormFactors}, making the $\kappa_4$ linear contribution about a 15\% of the $\kappa_4$-independent one.

\begin{figure}[htb]
    \centering
    \includegraphics[width = .75 \columnwidth]{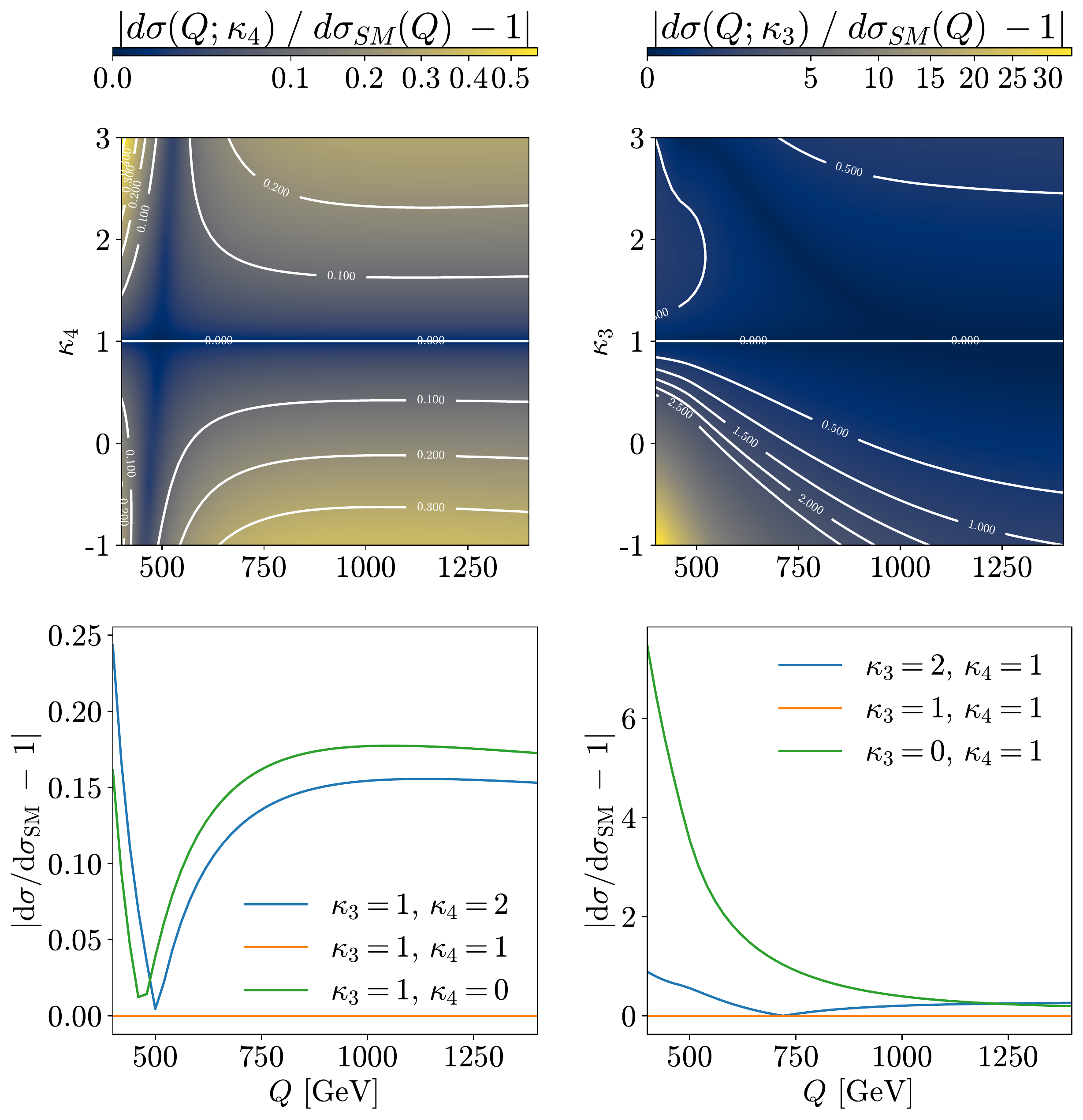}
    \caption{Departure from the SM of the triple Higgs boson invariant mass distribution for continuum (top) or discrete (bottom) values of $\kappa_4$ (left) and $\kappa_3$ (right). Due to the quadratic dependence of $\sigma$ on the couplings, a nonlinear colormap is used for better visualisation.}
    \label{fig:LO-diff-departure}
\end{figure}

Because of the different dependence on the self-couplings of the contributions presented in Eq.~\eqref{eq:LO-contributions}, a departure from the SM value $\kappa_{3,4} = 1$ might spoil the destructive interference patterns depicted in Figure~\ref{fig:LO-contributions}. This can be seen clearly in Figure~\ref{fig:LO-diff-departure},  where a large sensitivity to the $\kappa_3$ coupling is present particularly around the threshold, where the cross-section can be more than 30 times larger than the SM expectation. Indeed, deviations from $\kappa_3=1$ spoil the cancellation between the diagrams with most significant contributions, $\P$ and $\B$. In the case of $\kappa_4$, because it only affects the $2{\rm Re}((\P+\B)\T_4^*)$ contribution, which is suppressed with respect to $|\P+\B|^2$, the regions of larger sensitivity are those in which the latter is small, namely the production threshold (for $\kappa_3=1$) and the tail of the invariant mass distribution. 

\begin{figure}
    \centering
    \includegraphics[width=\columnwidth]{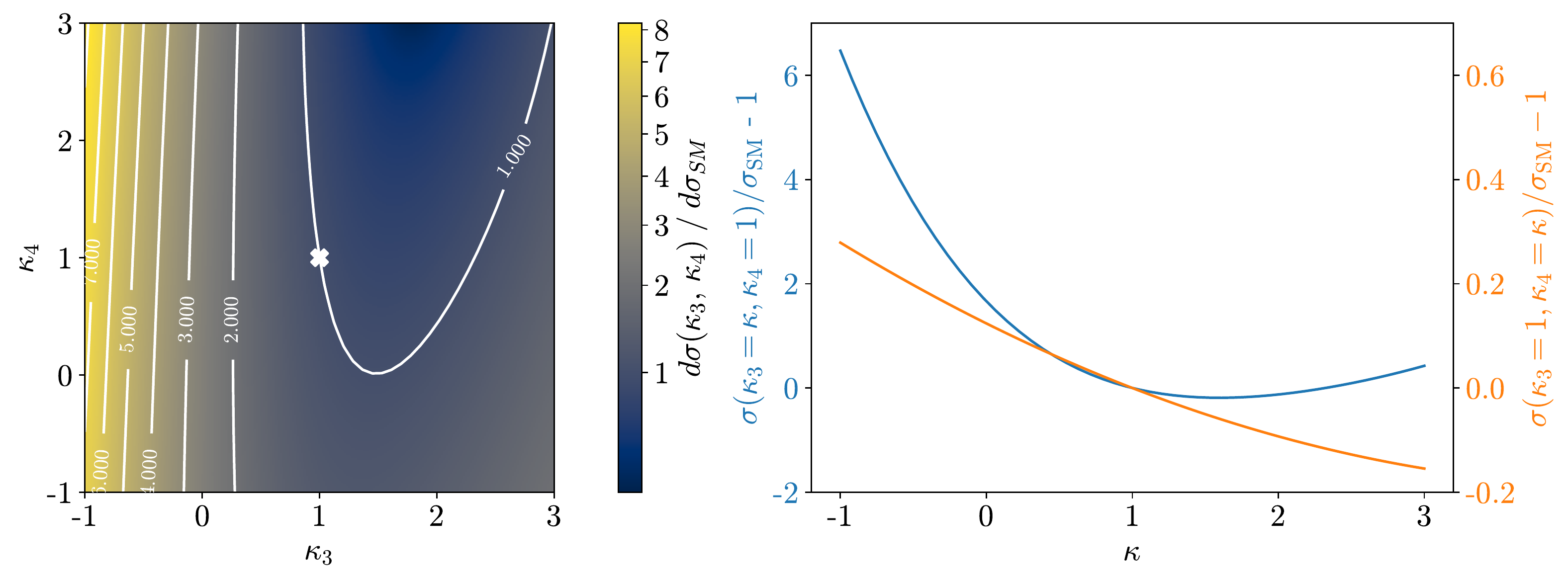}
    \caption{Departure of the inclusive cross-section as a function of $\kappa_3$ and $\kappa_4$ from its SM value. In the left plot the couplings are varied simultaneously and in the right one separately, keeping one fixed to the SM value. Due to the quadratic dependence of $\sigma$ on the couplings, a nonlinear colormap is used for better visualisation.}
    \label{fig:LO-tot-departure}
\end{figure}

We can also analyse the dependence of the inclusive cross-section on the couplings $\kappa_{3,4}$, by performing variations with respect to the SM value, as shown in Figure~\ref{fig:LO-tot-departure}.
For illustrative purposes, both coupling modifiers are varied in the range $\kappa_i \in [-1;3]$.
While the dependence on $\kappa_3$ is large (e.g. $\sigma > 8 \sigma_{\text{SM}}$ for $\kappa_3\sim -1$ and $\kappa_4 = 1$) and quadratic contributions become noticeable, the dependence on $\kappa_4$ is rather small (with departures of $|\sigma/\sigma_{\text{SM}} - 1| < 28\%$ in the range under study), which is compatible with the right panel of Figure~\ref{fig:LO-contributions} that shows the  $|\T_4|^2$ contribution to the invariant mass distribution.

\section{NNLO Corrections}
\label{sec:correction}
After discussing the different contributions to the LO cross-section in the previous section, we will present the results for the full NNLO corrections in the HTL.

In the HTL the Higgs bosons couple directly to gluons via the effective Lagrangian
\begin{equation}
    \mathcal{L}_{\rm eff} = - \frac{1}{4} G^a_{\mu \nu} G_a^{\mu \nu} \left(C_H \frac{H}{v} - C_{HH} \frac{H^2}{2 v^2} + C_{HHH} \frac{H^3}{3 v^3} + \ldots\right),
\end{equation}
where the matching coefficients can be expanded in powers of the strong coupling constant $\as$ as
\begin{equation}
    C_X = - \frac{\as}{3 \pi} \sum_{n\ge0} C_X^{(n)} \left(\frac{\as}{\pi}\right)^n,
\end{equation}
where the expansion is known up to fourth order for $X= H, \, HH,\, HHH$\cite{Chetyrkin:1997iv,Chetyrkin:2005ia,Kramer:1996iq,Schroder:2005hy,Djouadi:1991tka,Grigo:2014jma, Spira:2016zna}. 

The computation of the complete NLO corrections and of the virtual amplitudes up to NNLO, both in the HTL, was presented in Ref.~\cite{deFlorian:2016sit}. The latter were used to construct the soft-virtual approximation (based on the results from Ref.~\cite{deFlorian:2012za}) to provide a phenomenological NNLO$_\text{SV}$ cross-section. In the present work we computed the real emission amplitudes to obtain a full NNLO cross-section in the HTL, including all partonic channels in addition to gluon-fusion. To this end, we exploited the known relation between the single Higgs boson cross-section and some contributions to the multiple Higgs one. This is done in the same fashion as the calculations for double Higgs production \cite{deFlorian:2013jea} and its extension to the dimension 6 Standard Model Effective Theory and Higgs Effective Theory \cite{deFlorian:2017qfk}. 

Contributions involving only one HTL operator at the amplitude level (i.e. a single effective vertex between Higgs bosons and gluons) are totally equivalent, apart for an overall normalization and the corresponding matching coefficient, to those from single Higgs production.
In terms of the degree of difficulty, these contributions are {\it truly} at the NNLO level, but the results can be borrowed from single Higgs production, exploiting the above mentioned similarities.
On the other hand, when considering diagrams with more than one HTL operator insertion (actually, their interference with the ones with just one insertion), as each of them carries an extra coupling $C_X = \mathcal{O}(\as)$, only tree level configurations can appear to NLO, while at NNLO accuracy the  only possible contributions arise from one-loop and single real emission diagrams. Because of this simplification, their infrared divergences can actually be handled with standard NLO procedures.

In our calculation, the {\sc FeynArts}\cite{Hahn:2000kx} and {\sc FeynCalc}\cite{Mertig:1990an,Shtabovenko:2016sxi} packages were used in {\sc Mathematica} to compute the amplitudes. The cancellation of their infrared singularities with the ones present in the virtual amplitudes presented in Ref.~\cite{deFlorian:2016sit} and the absorption of the remaining ones into the evolution of the parton distribution functions was performed following the FKS\cite{Frixione:1995ms,Frixione:1997np} approach. In this work we neglect effects coming from the Higgs width, and set it equal to zero throughout the calculation. 

We present below the final result with a notation suitable for the discussion on the reweighting following in Section~\ref{sec:Reweight}. More details on the derivation can be found in the Appendix.

As usual, the cross-section can be written as 
\begin{equation}
    \frac{\dif{\sigma}}{\dif{Q^2}} = \sum_{i,j} \int_0^1 \dif{x_1}\dif{x_2} f_{i/h_1}(x_1) f_{j/h_2}(x_2) \int_0^1 \delta\left(x - \frac{Q^2}{x_1 x_2 s_H}\right) \frac{\dif{\hat\sigma_{i j}}}{\dif{Q^2}},
\end{equation}
where $Q$ is the invariant mass of the triple Higgs system, $\sqrt{s_H}$ the collider centre of mass energy, and $i,j$ are the labels for the massless partons inside the hadrons $h_1$ and $h_2$ with respective parton density $f_{(i,j)/(h_1,h_2)}$. Here the dependence on the factorisation and the renormalisation scales is implicitly understood.

The partonic cross-section $\hat \sigma$ is computed order by order as an expansion in the strong coupling $\alpha_S$, such that up to NNLO we write it as

\begin{equation}
    Q^2 \frac{\dif{\hat \sigma_{ij}}}{\dif{Q^2}} = \frac{1}{2 Q^2 3!\, 2^2} \int \dif{\rm PS}_3 |\M_{3H}|^2 \left[ \eta_{ij}^{(0)} + \astwopi \eta_{ij}^{(1)} + \astwopi^2 \eta_{ij}^{(2)} + \O(\as^3) \right],
    \label{eq:part-sigma}
\end{equation}
where the LO amplitude $\M_{3H}$
can generically be written as
\begin{equation}
    \M_{3 H} = \astwopi \frac{Q^2}{3 v^3} C_\text{LO}^{3H},
\end{equation}
and
\begin{align}
    C_\text{LO}^{3H}(p_1,p_2,p_3,p_4,p_5) =& F_\P(p_1,p_2,p_3,p_4,p_5) +F_\T (Q^2) \, \frac{3 m_H^2 \kappa_4}{Q^2 - m_H^2} + \sum_{(kl)} \Big[\frac{(3 m_H^2 \kappa_3)^2}{Q^2 - m_H^2} \,F_\T(Q^2) \nonumber \\+&  3 m_H^2 \kappa_3 \,F_\B(Q^2,(p_k+p_l-p_1)^2,(p_k+p_l-p_2)^2,s_{kl}, m_H^2)\Big]\frac{1}{s_{kl} - m_H^2} .
\end{align}

The perturbative coefficients $\eta^{(0,1,2)}_{ij}$, within the HTL approximation and its extensions (reweightings) described in Section~\ref{sec:Reweight}, are up to NNLO
\begin{align}
    \eta_{ij}^{(0)} =&\, \eta_{ij}^{H\,(0)} = \delta(1-x)\delta_{ig}\delta_{jg}, \\
    \eta_{ij}^{(1)} =&\, 2 \,\eta_{ij}^{H\,(1)} + \eta_{ij}^{H\,(0)} \,\frac{4}{3} \,\frac{{\rm Re}\left( (C_\text{LO}^{3H})^* \sum_{(kl)} C_\text{LO}^{2H}(\{k,l\}) \right)}{|C_\text{LO}^{3H}|^2},\\
    \eta_{ij}^{(2)} =&\,  4\, \eta_{ij}^{H\,(2)} + 4 \eta_{ij}^{H\,(0)} \frac{2\,{\rm Re}\left( (C_\text{LO}^{3H})^* \left( (C_{HHH}^{(2)}-C_{H}^{(2)}) \P + (C_{HH}^{(2)}-C_{H}^{(2)})\kappa_3 \B\right)\right)}{|C_\text{LO}^{3H}|^2}\nonumber \\
    &+ \eta_{ij}^{H\,(0)} \frac{{\rm Re}\left((C_\text{LO}^{3H})^* (\R^{(2)}_{3H} + (C_\text{LO}^{3H})^3 \,\T^{(2)}_{3H})\right) + \V^{(2)}_{3H} |C_\text{LO}^{H}|^2}{|C_\text{LO}^{3H}|^2} + \rho^{(2)}_{ij} ,
\end{align}
where $\eta^{H,(n)}_{ij}$, $n=0,1,2$, are the corresponding QCD corrections for single Higgs production that can be found in Ref.~\cite{Anastasiou:2002yz} (and coincide with the results in Refs.~\cite{Harlander:2002wh,Ravindran:2003um}), the renormalisation and factorisation scales were set to $\mu_F = \mu_R = Q$, $k$ and $l$ label the final state Higgs bosons, and $\sum_{(kl)}$ denotes the sum over distinct pairs of them. $C_\text{LO}^{H}$ and $C_\text{LO}^{2H}$ are (up to a normalisation factor) the LO amplitudes for single and double Higgs production
\begin{align}
    C_\text{LO}^{H} =& F_\T(m_H^2),\label{eq:CLO_H}\\
    C_\text{LO}^{2H}(\{k,l\}) =& \frac{3 m_H^2}{s_{kl}-m_H^2} F_\T(s_{kl}) + F_\B(\{k,l\}),\label{eq:CLO_2H}
\end{align}
and the coefficients $\R_{3H}^{(2)}$, $\T_{3H}^{(2)}$ and $\V_{3H}^{(2)}$ are the finite remainders of the virtual corrections at NNLO presented in Ref.~\cite{deFlorian:2016sit}, with the only difference being that we use (as we will discuss later) the general definition of Eq.~\eqref{eq:LO-FormFactors} for $C_\text{LO}^{2H}$. If we express this coefficient in the HTL, the expressions are identical to those in Ref.~\cite{deFlorian:2016sit}.

The only missing ingredient, $\rho_{ij}^{(2)}$, corresponds to the finite remainder of the real emission corrections to the diagrams with more than one HTL operator insertion (which are already included in $\eta_{ij}^{H(2)}$). These only contain diagrams with a single parton emission whose divergences were regulated using dimensional regularisation in $D=4-2\epsilon$ dimensions in the FKS\cite{Frixione:1995ms,Frixione:1997np} framework. After subtracting the singularities, we can write the remainder as
\begin{align}\label{eq:reals}
 \rho^{(2)}_{ij} &= \frac{4}{3} \,\frac{{\rm Re}\left( (C_\text{LO}^{3H})^* \sum_{(kl)} C_\text{LO}^{2H}(\{k,l\}) \right)}{|C_\text{LO}^{3H}|^2} \rho_{ij}^{(sc)} \nonumber \\
 &+\int_0^{2\pi} \dif{\phi} \int_{-1}^1 \dif{y} \frac{1}{2}\left(\frac{1}{1-x}\right)_+ \left[\left(\frac{1}{1-y}\right)_+ + \left(\frac{1}{1+y}\right)_+\right] \rho^{(r)}_{ij} (x, y, \phi)
\end{align}
that contains a soft-collinear term $\rho_{ij}^{(sc)}$ and a regular term $\rho^{(r)}_{ij}$ whose explicit expressions, together with the definition of the standard plus-prescription, are given in the Appendix.

\subsection{Reweighting}
\label{sec:Reweight}
The results presented in this section complete the full NNLO corrections to triple Higgs production in the HTL when the quantities $F_\P$, $F_\B$ and $F_\T$ are expressed in this limit (see Eq.~\eqref{eq:LO-FormFactors}). In order to retain part of the $m_t$ dependence, this expression is usually reweighted by using the exact result for $\M_{3H}$ in Eq.~\eqref{eq:part-sigma}, while keeping the coefficients $\eta^{(0,1,2)}_{ij}$ in the HTL. This procedure is usually referred to as the \emph{Born-improved} (Bi) NNLO cross-section, and its accuracy has been studied for double Higgs production~\cite{Borowka:2016ehy,Borowka:2016ypz,Baglio:2018lrj} at NLO finding that it overestimates the exact inclusive cross-section by a 32\% at collider energies of 100~TeV (and around 16\% at 14~TeV).
This overestimation is enhanced in the tail of the Higgs pair invariant mass distribution.

In order to parametrise the dependence of the result on the reweighting procedure, we also considered what we call a \emph{dynamically-Born-improved} (dBi) NNLO cross-section, which we obtain by using the full dependence on the kinematics of the outgoing particles of 
$F_\T$, $F_\B$ and $F_\P$ in the definitions of $\P$, $\B$, $C_\text{LO}^{H}$ and $C_\text{LO}^{2H}$, and therefore also in the definitions of $\R^{(2)}_{3H}$,$\T^{(2)}_{3H}$, $\V^{(2)}_{3H}$ and $\rho_{ij}^{(2)}$. In this way, we reweight the HTL insertion operators with the respective LO amplitude diagram by diagram (e.g. the $(H)Hgg$ vertex with the (double) single Higgs production amplitude $\sim C_\text{LO}^{(2)H}$).
This reweight cannot be applied in a straightforward way, as the form factor $F_\B$ is not always defined for the kinematics characterising the HTL vertex. This happens in diagrams with two HTL vertices connected via an off-shell gluon (e.g. a box and a triangle loop). 
To fix this problem, we modify the kinematics in a way such that we preserve the momenta of the outgoing Higgs bosons, while redefining the momenta of the gluons entering the vertex to be on-shell. 

For the triangle form factor $F_\T$, we evaluate it at the invariant mass of the outgoing Higgs boson, just as expressed in the Eqs.~\eqref{eq:CLO_H} and \eqref{eq:CLO_2H}. For the box form factor $F_\B$ appearing in Eq.~\eqref{eq:CLO_2H}, we need to define a prescription that corrects the momenta of the initial gluons to compensate for the recoil of all other particles not involved in the HTL vertex. 
Lets recall we are labelling $p_1$ and $p_2$ as the momenta of the incoming partons, and $p_{(3,4,5)}$ the momenta of the outgoing Higgs Bosons. For the vertex with outgoing Higgses $\{i,j\}$, we define $q := p_i + p_j$, $M^2 := q^2$, and $q_T^\mu$ as the transverse component  of $q$ with respect to $p_1$ and $p_2$, and then define the momenta of the gluons entering the form factor as $k_1$ and $k_2 := q - k_1$ in the following way
\begin{align}
    F_\B (\{i,j\}) =& F_\B ((k_1+k_2)^2, (k_1-p_i)^2, (k_1-p_j)^2, m_H^2, m_H^2),\\
    k^\mu_1 :=& z_1 \frac{M^2}{2 q \cdot p_1} p_1^\mu + \xi q_T^\mu + \frac{\xi \pmb{q_{T}}^2}{z_1} \frac{q\cdot p_1}{M^2 p_1 \cdot p_2}p_2^\mu,\qquad (q_T^\mu q_{T\mu} =: -\pmb{q_T}^2),\\
    z_1 :=& \frac{M^2 + 2 \xi \pmb{q_T}^2 + \sqrt{(M^2 + 2 \xi \pmb{q_T}^2)^2 - 4 (M^2 + \pmb{q_T}^2) \xi^2 \pmb{q_T}^2}}{2 M^2},
\end{align}
where the different choices of $\xi\in[0,1/2]$ define different consistent prescriptions to account for the Higgs pair recoil. This prescription corresponds to the one presented in Ref.~\cite{Catani:2015vma} in the context of transverse-momentum resummation, if one chooses $\pmb{k_{1T}} = \xi \pmb{q_T}$ for the Eqs.~(25) and (26) therein. In particular, if $\xi=0$ the transverse recoil is compensated by one gluon ($\pmb{k_{1T}} = 0$, $\pmb{k_{2T}} = \pmb{q_T}$) while if $\xi = 1/2$ it is equally compensated by both gluons ($\pmb{k_{1T}} = \pmb{k_{2T}} = \pmb{q_T}/2$).

Let us emphasise that this prescription leaves unchanged the momenta of all the Higgs bosons involved, including their virtuality, and its main purpose is to redefine the momenta of the virtual gluons entering Box diagrams, so that these are on-shell and $F_\B$ is well defined.

With this prescription, the \emph{dynamically Born improved} approximation is well defined and we will denote it as dBi$_\xi$. In particular, we will present results for dBi$_0$ and dBi$_{1/2}$ as benchmarks to show the dependence on the choice of $\xi$, and we will show that this dependence is numerically negligible at NLO and NNLO.

\section{Phenomenological Results}
\label{sec:pheno}
We present results for the NNLO cross-section for triple Higgs production, reweighted using the Bi, dBi$_0$ and dBi$_{1/2}$ prescriptions. The set-up is the same as the one used in Section~\ref{sec:LO}. 
We use the MMHT2014\cite{Harland-Lang:2014zoa} set of parton distributions, at the corresponding order in the strong coupling constant for each contribution (LO, NLO, NNLO). 
For phenomenological purposes, we compute this for collider energies of 100~TeV and 27~TeV that are relevant for physics at the Future Circular Collider and High-Energy LHC, respectively. To estimate the theoretical uncertainty arising from the missing higher orders in the perturbative series, we perform an independent variation of the factorisation and renormalisation scales in the range $[\mu_0 / 2, 2\mu_0]$, with the constrain $0.5<\mu_R/\mu_F < 2$. The choice of the central values $\mu_0$ used in this work were $\mu_0 = Q/2$ and $\mu_0 = Q$, where $Q$ is the invariant mass of the  triple Higgs system. 

\begin{figure}[ht]
    \centering
    \includegraphics[width=\columnwidth]{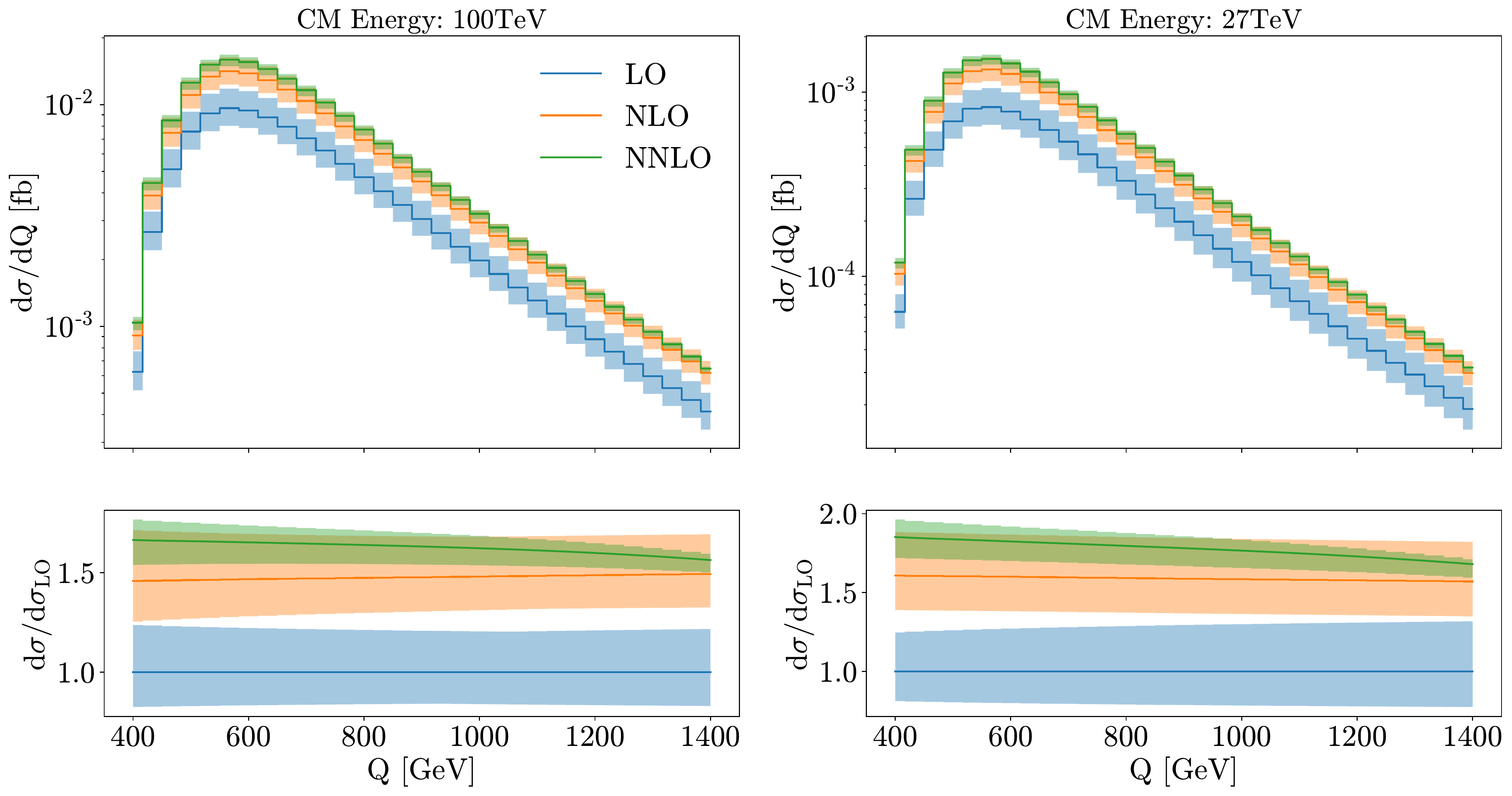}
    \caption{Invariant mass distribution of the  triple Higgs system in the dBi approximation up to different orders (up) and corresponding $K$ factor (down). The results are shown for a collider center of mass energy of 100~TeV (left) and 27~TeV (right). The shaded bands correspond to the uncertainty from the variation of scales from the central value of $\mu_0 = Q/2$.}
    \label{fig:pheno_kfactor}
\end{figure}

In Figure~\ref{fig:pheno_kfactor} we see the cross-section computed in the dinamically Born improved approximation up to different orders, as well as the $K$ factor defined as usual, $K = \dif{\sigma} / \dif{\sigma_\text{LO}}$.  As seen also within the soft-virtual approximation~\cite{deFlorian:2016sit}, the cross-section begins to stabilise only from NNLO. The $K$ factors are rather flat at the peak of the invariant mass distribution, with values around 1.7 and 1.8 for collider CM energies of 100 and 27~TeV respectively, while the NNLO $K$ factors present a suppression in the tail. Due to this suppression, the entire NNLO band falls inside the scale variation of the NLO, suggesting that the perturbative series is more stable in this region. 
The total scale uncertainty is reduced from  37\% to 27\% and to 11\% when going from LO to NLO to NNLO, at 100~TeV with a central scale choice of $\mu_0 = Q/2$. For 27~TeV the reduction is similar, from 48\% to 30\% and then to 12\%.

\begin{figure}[tb]
    \centering
    \includegraphics[width=\columnwidth]{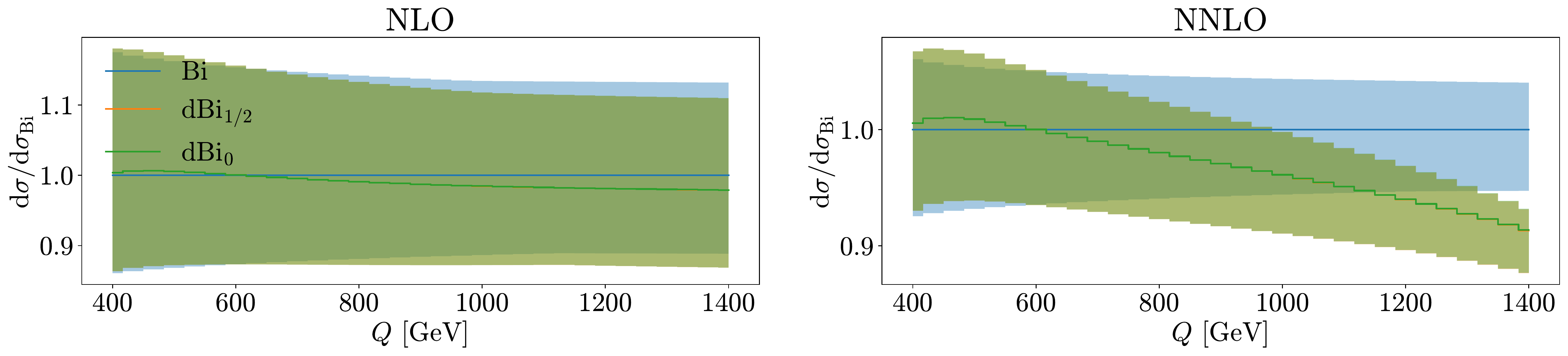}
    \caption{
    Comparison between the Bi, dBi$_0$ and dBi$_{1/2}$ reweights to the triple Higgs boson invariant mass distribution. The result is shown at NLO (left) and NNLO (right) in the $\alpha_S$ expansion. There are no visible differences between the dBi$_0$ and dBi$_{1/2}$ bands, as they overlap completely.
    The shaded bands correspond to the uncertainty from the variation of scales from the central value of $\mu_0 = Q/2$.}
    \label{fig:pheno_comparison}
\end{figure}

To measure the effect of the reweighting, we can compare the different approximations Bi, dBi$_0$ and dBi$_{1/2}$ and observe the corresponding effect on the invariant mass distribution. In Figure~\ref{fig:pheno_comparison} we see that, although at NLO the three approximations are almost completely compatible, at NNLO there is a significant decrease in the tail of the distribution when using a dBi$_\xi$ instead of the Bi approximation, a discrepancy that is even bigger than the scale variation in this region. We also notice that the dependence on $\xi$ of the dBi$_\xi$ approximation is phenomenologically negligible. 
We know that for double Higgs production the Bi approximation overestimates the tail of the distribution at NLO respect to the exact calculation~\cite{Borowka:2016ehy,Borowka:2016ypz,Baglio:2018lrj}. If this effect holds also for triple Higgs production, we can expect the dBi approximation to provide more reliable predictions, as it predicts a smaller tail of the distribution just by reweighting each contribution by the associated form factor instead of using the full amplitude.

In order to understand why the discrepancies in the tail of the invariant mass distribution between the Bi and dBi prescriptions arise only at NNLO, lets recall what are the main differences between the two reweighting procedures. The Bi reweights all amplitudes by the Born amplitude, including those that have more than one HTL vertex. The dBi only does this to amplitudes containing a single HTL vertex, while applying a different prescription for those amplitudes with more than one HTL vertex. In this way, the Bi reweighting procedure increases the relative significance of the amplitudes with many HTL vertices, respect to the dBi. At NLO such amplitudes appear only at tree level, due to the power counting of the HTL vertices, making the discrepancy phenomenologically negligible. At NNLO, such diagrams are enhanced by real emission corrections, making the discrepancy at the tail of the invariant mass distribution more noticeable. 

\begin{table}[ht]
    \centering
    \renewcommand{\arraystretch}{1.2}
    \begin{tabular}{l|c c c }
        \hline\hline
        $\mu_0 = Q$ & 14~TeV & 27~TeV & 100~TeV  \\
        \hline
        LO & $0.0462^{+31\%}_{-22\%}$ & $0.235^{+26\%}_{-19\%}$& $3.29^{+20\%}_{-15\%}$ \\[0.6ex]
        \hline
        NLO$_{\rm Bi}$ & $0.0833^{+18\%}_{-15\%}$  & $0.408^{+16\%}_{-13\%}$ & $5.12^{+14\%}_{-11\%}$ \\[0.6ex]
        NLO$_{\rm dBi}$ & $0.0831^{+18\%}_{-15\%}$ & $0.407^{+16\%}_{-13\%}$& $5.09^{+14\%}_{-12\%}$ \\[0.6ex]
        \hline
        NNLO$_{\rm Bi}$ & $0.105^{+8\%}_{-9\%}$ & $0.503^{+7\%}_{-8\%}$& $6.11^{+6\%}_{-7\%}$ \\[0.6ex]
        NNLO$_{\rm dBi}$ & $0.104^{+8\%}_{-9\%}$ & $0.498^{+7\%}_{-8\%}$& $6.02^{+6\%}_{-7\%}$ \\[0.6ex]
        \hline\hline
        $\mu_0 = Q/2$ & 14~TeV & 27~TeV & 100~TeV  \\
        \hline
        LO & $0.0605^{+34\%}_{-24\%}$ & $0.295^{+28\%}_{-20\%}$& $3.88^{+21\%}_{-16\%}$ \\[0.6ex]
        \hline
        NLO$_{\rm Bi}$ & $0.0983^{+18\%}_{-15\%}$  & $0.473^{+16\%}_{-14\%}$ & $5.75^{+15\%}_{-12\%}$ \\[0.6ex]
        NLO$_{\rm dBi}$ & $0.0982^{+18\%}_{-15\%}$ & $0.471^{+17\%}_{-14\%}$& $5.72^{+15\%}_{-12\%}$ \\[0.6ex]
        \hline
        NNLO$_{\rm Bi}$ & $0.114^{+5\%}_{-8\%}$ & $0.540^{+5\%}_{-7\%}$& $6.47^{+5\%}_{-6\%}$ \\[0.6ex]
        NNLO$_{\rm dBi}$ & $0.113^{+5\%}_{-8\%}$ & $0.534^{+5\%}_{-7\%}$& $6.36^{+5\%}_{-6\%}$ \\[0.6ex]
        \hline\hline
        NNLO$_\text{Best}$ & $0.103^{+5\%}_{-8\%}$ & $0.501^{+5\%}_{-7\%}$& $5.56^{+5\%}_{-6\%}$ \\[0.6ex]
        \hline\hline
    \end{tabular}
    \caption{Results for the inclusive cross-section (in fb) of triple Higgs boson production for different collider energies, calculated at different orders and with the different reweighting procedures.
    The results are shown for central scale values of $Q$ (top) and $Q/2$ (bottom). The dependence on $\xi$ in the dBi$_\xi$ reweight procedure is below the per-mill level and therefore omitted.
    The last row shows our best available prediction for the different collider energies.
    The uncertainties correspond to the scale variation.}
    \label{tab:Cross_sections}
\end{table}

In Table~\ref{tab:Cross_sections} we present the results for the inclusive cross-section obtained for different collider energies, choices of central scales $\mu_0$, reweighting procedures and orders in the perturbative expansion. 
The corresponding $K$ factors for the dBi results are presented in Table~\ref{tab:K_factors}.
When comparing our results with the ones obtained in Ref.~\cite{deFlorian:2016sit} in the soft-virtual approximation, we find that although the SV result differs only in about 1\% from the full NNLO in the HTL, when using the Bi reweight the difference grows to a 2.5\% and 4.4\% increase in the inclusive cross-section at 14~TeV and 100~TeV, respectively.
This larger difference is due to the fact that in the HTL the SV approximation is slightly smaller than the complete NNLO for small invariant masses, but compensates at large invariant masses, resulting in a small difference in the inclusive cross section. After the reweighting procedure, the region of large invariant masses is suppressed, and therefore this accidental compensation is reduced, increasing the difference in the inclusive cross sections between SV and complete NNLO.

\begin{table}[htb]
    \centering
    \renewcommand{\arraystretch}{1.2}
    \begin{tabular}{l|c c c }
        \hline\hline
        $\mu_0 = Q$ & 14~TeV & 27~TeV & 100~TeV  \\
        \hline
        $K_{\rm NLO}$ &1.80 & 1.732 & 1.56\\[0.6ex]
        \hline
        $K_{\rm NNLO}$ &2.27 & 2.12 &1.55 \\[0.6ex]
        \hline\hline
        $\mu_0 = Q/2$ & 14~TeV & 27~TeV & 100~TeV  \\
        \hline
        $K_{\rm NLO}$ &1.63 &1.60  &1.47 \\[0.6ex]
        \hline
        $K_{\rm NNLO}$ &1.87 &1.81 & 1.64\\[0.6ex]
        \hline\hline
        $K_{\rm NNLO-Best}$ &1.70 &1.70 &1.43\\[0.6ex]
        \hline\hline
    \end{tabular}
    \caption{NLO and NNLO $K$ factors for the inclusive triple Higgs boson production at different collider energies. These defined as the quotient between the dBi and the LO cross section. 
    The results are shown for central scale values of $Q$ (top) and $Q/2$ (bottom). The dependence on $\xi$ in the dBi$_\xi$ reweight procedure is below the per-mill level and therefore omitted. The last row shows our best available prediction for the different collider energies.}
    \label{tab:K_factors}
\end{table}

 The different reweighting procedures produce a small difference ranging from $\sim 0.7\%$  at 14~TeV up to $\sim 1.3\%$ at 100~TeV. Although in Figure~\ref{fig:pheno_comparison} we see that the different reweights lead to discrepancies larger than the theoretical uncertainties in the tail of the distribution, since this region has a small impact on the inclusive cross-section the results for the total cross-section are consistent within the theoretical uncertainties. Of course, this small difference at the total cross section level can only be a lower bound on the expected finite top mass effects, and from the results obtained at NLO within the FT$_\text{approx}$ (which are $\sim 10\%$ smaller than the dBi prediction) it is clear that they are expected to be much larger. What we can conclude from this exercise therefore, is that the systematic uncertainties related to the choice of the reweighting procedure (among the choices presented here) is expected to be marginal compared to the full size of the finite-$m_t$ effects.

In order to provide the best possible estimate of the triple-Higgs production cross section, it becomes necessary to include the partial finite-$m_t$ effects obtained in Ref.~\cite{Maltoni:2014eza} within the FT$_\text{approx}$. To this end, we use the predictions presented therein and in Ref.~\cite{deFlorian:2016spz} for the total cross section, and encode the finite mass effects in the parameter $\delta_t$ defined by
\begin{equation}
\sigma^\text{NLO}_\text{FTapprox} = \sigma^\text{NLO}_\text{dBi} (1+\delta_t)\,,
\end{equation}
 and we define our best prediction as
\begin{equation}
\sigma^\text{NNLO}_\text{Best} = \sigma^\text{NNLO}_\text{dBi}  + \delta_t \sigma^\text{NLO}_\text{dBi} \,. 
\end{equation}
 This procedure is similar to the prescription that was implemented in Ref.~\cite{deFlorian:2016spz} for double-Higgs production.
 The values that we obtain for $\delta_t$ at the different collider energies are $\delta_t =$ -0.107, -0.073 and -0.146 for 14, 27 and 100~TeV, respectively.\footnote{
 The value corresponding to 27~TeV was extracted from results of~\cite{Maltoni:2014eza} computed at 33~TeV, which is the closest one in energy available in the literature.
} The corresponding cross sections and K factors are presented in Tables \ref{tab:Cross_sections} and \ref{tab:K_factors}, respectively.

 Before providing our final results, we address the issue of the remaining uncertainties associated to finite-$m_t$ effects.
 There is in principle no reason to expect these effects to be smaller than the corresponding ones present in double Higgs production; in fact the typically larger invariant masses involved might point to the opposite direction. Of course, the more complicated structure of the triple Higgs production amplitudes, involving additional topologies, might lead to accidental cancellations of these effects that cannot be predicted at this point.
 In the absence of any prediction with full top mass dependence beyond the LO, one can only rely on approximated results in order to estimate the associated uncertainties at NNLO. The best available approximation at NLO, the FT$_\text{approx}$, differs from the Born-improved NLO at ${\cal O}(10\%)$ at 14 and 27~TeV, and ${\cal O}(15\%)$ at 100~TeV. The size of this difference can be a good estimation of the missing finite top mass effects, and indeed this is the case for the double-Higgs production cross section. In order to provide a conservative estimation, and having in mind the possibly worse situation in triple Higgs as compared to Higgs pair production, we estimate the uncertainty of our prediction to be of $\pm 15\%$ at 14 and 27~TeV, and $\pm 20\%$ at 100~TeV.
 
 Given that the full dependence on the top mass is only retained at LO, it is not possible to perform a complete analysis on the top mass scheme uncertainty (which was found to be large at NLO in the case of double Higgs~\cite{Baglio:2018lrj}). Nevertheless, a parametric variation of the default on-shell value $m_t=173.2$~GeV used along this work  to the correspondent one in the $\overline{\rm MS}$ scheme, $\overline{m}_t(\overline{m}_t)=163.6$~GeV (using a three-loop conversion between schemes), shows a decrease of about 25\% in the cross section, which indicates an uncertainty in line with the one estimated for the finite top mass effects.

 Compiling all the ingredients described in the last paragraphs, we arrive therefore to the following final prediction for the triple Higgs production cross section:
\begin{align*}
\sigma^\text{NNLO}_\text{Best} &= 0.103^{+5\%}_{-8\%}  \pm 15\% {\rm \, fb},\hspace{1cm}
    K^\text{NNLO}_\text{Best} = 1.70, \hspace{1cm} (14\text{ TeV})
\\[1ex]
\sigma^\text{NNLO}_\text{Best} &= 0.501^{+5\%}_{-7\%}  \pm 15\% {\rm \, fb},\hspace{1cm}
    K^\text{NNLO}_\text{Best} = 1.70, \hspace{1cm} (27\text{ TeV})
\\[1ex]
\sigma^\text{NNLO}_\text{Best} &= 5.56^{+5\%}_{-6\%}  \pm 20\% {\rm \, fb},\hspace{1.2cm}
    K^\text{NNLO}_\text{Best} = 1.43. \hspace{1cm} (100\text{ TeV})
\end{align*}

\section{Conclusions}
\label{sec:conc}
In this work we presented for the first time the complete set of NNLO corrections to triple Higgs boson production at hadron colliders in the heavy top limit. To partially retain finite top mass effects, two different reweighting procedures have been implemented: The usual Born-improved approximation (Bi), and a new procedure that we call dynamically Born improved approximation (dBi). Both procedures coincide at LO, and from their difference at higher orders we infer the dependence of the result upon the reweighting procedure. 
Overall, we found that the invariant mass distribution is sensitive to the reweighting procedure only in the tail, where the cross-section is already small, while for the inclusive cross-section the dependence on this procedure is $\mathcal{O}(1\%)$, which falls inside the scale variation uncertainties of $\mathcal{O}(7\%)$.

In order to provide a prediction based on the most advanced results available in the literature, we combined our dBi-reweighted NNLO results with the NLO predictions obtained within the FT$_\text{approx}$. From the differences between the available NLO approximations we estimated the size of the missing finite top mass effects. Based on this, our final prediction for the triple Higgs production cross section at a 100~TeV collider is $\sigma_\text{NNLO}=5.56^{+5\%}_{-6\%}  \pm 20\%$~fb.

\section*{Acknowledgements}

This work is supported in part by the Swiss National Science Foundation (SNF) under contracts 200020\_169041 and IZSAZ2\_173357, by MINCyT under contract SUIZ/17/05, by Conicet and by ANPCyT.
We thank Jean-Nicolas Lang for his support with the {\sc Recola2} library.

\bibliography{biblio}

\newpage
\appendixtitleon
\appendixtitletocon
\begin{appendices}

\allowdisplaybreaks

\section*{Appendix: Real corrections}
\renewcommand{\theequation}{{\rm{A}}.\arabic{equation}}
\setcounter{equation}{0}

In this appendix we  present the results for the corrections arising from real emission in diagrams with more than one HTL insertion operator. 

Because each HTL operator insertion carries a $\as$ factor, tree level diagrams with two operator insertions are already NLO, so their real radiation corrections correspond to a single parton emission. In the same way, diagrams with three operator insertions appear only at NNLO and their real emission corrections are of higher order, and therefore not considered. The single real emission amplitudes present divergences when the emitted parton becomes unresolved, some of which will cancel against the divergences present in the corresponding loop corrections calculated in Ref.~\cite{deFlorian:2016sit} and the rest have to be absorbed in the NLO evolution of the parton distribution functions. In order to perform such cancellations, we used an FKS approach in $D=4-2\epsilon$ dimensions. 

The key idea of the FKS method is to divide the phase space of the real corrections ${\rm dPS}_{4}$ into soft, collinear and regular regions. To do so, we express it in terms of the LO phase space ${\rm dPS}_3$ plus the dependence on the emitted particle
\begin{align}\label{eq:ap_dps}
    {\rm dPS}_{4} \left|\M_r\right|^2=& (4\pi)^{-2+\epsilon} \frac{\Gamma(1-\epsilon)}{\Gamma(1-2\epsilon)} \frac{ s^{1-\epsilon}}{2 \pi} (1-x)^{-1-2\epsilon} (1-y^2)^{-1-\epsilon} \dif{y} \sin^{-2\epsilon}\phi \dif{\phi}  \nonumber \\& \left[(1-x)^2 (1-y^2) \left|\M_r\right|^2\right]{\rm dPS}^{(x)}_{3}
\end{align}
where $\M_{r}$ is the amplitude for the real emission process we are considering, $s$ is the invariant mass of the incoming partons, ${\rm dPS}^{(x)}_3$ is the Born phase space evaluated at $s \to x s$ and $Q^2$ is the invariant mass of the triple Higgs system such as $x = Q^2/s$. The emitted parton therefore has a momentum fraction of  $(1-x) s$, and we define its orientation with respect to one of the initial partons with the azimuth $\phi$ and the cosine of the polar angle $y$.

With this definitions, the real radiation amplitude $\M_r$ becomes singular in the regions $x\to 1$ (soft) and $y\to \pm1$ (collinear). Nevertheless, the expression $(1-x)^2 (1-y^2) \left|\M_r\right|^2$ is regular in both limits. This means that in Eq.~\eqref{eq:ap_dps} we have isolated the singularities in the terms $(1-x)^{-1-2\epsilon}$ and  $(1-y^2)^{-1-\epsilon}$. To make them explicit, we use the following identities
\begin{align}
    (1-x)^{-1-2\epsilon} =& \frac{-1}{2 \epsilon} \delta(1-x) + \left(\frac{1}{1-x}\right)_+ - 2 \epsilon \left(\frac{\log(1-x)}{1-x}\right)_+ + \mathcal{O}(\epsilon^2)\\
    (1-y^2)^{-1-\epsilon} =& -\frac{4^{-\epsilon}}{2 \epsilon} \left[\delta(1-y)+\delta(1+y)\right] + \frac{1}{2} \left[\left(\frac{1}{1-y}\right)_+ + \left(\frac{1}{1+y}\right)_+\right] + \mathcal{O}(\epsilon)
\end{align}
where the plus distributions are defined as usual
\begin{align}
    \int_0^1\dif{x} f(x) \left[g(x)\right]_+ =& \int_0^1\dif{x} (f(x) - f(1)) \, g(x)\\
    \int_{-1}^1\dif{y} f(y) \left(\frac{1}{1\pm y}\right)_+ =& \int_{-1}^1\dif{y} \frac{f(y) - f(\mp 1)}{1\pm y}.
\end{align}
Now that the different divergences are explicitly expressed as poles in $\epsilon$, we can subtract the infrared divergences appearing in the virtual amplitudes, and reabsorb the remaining  divergences in the factorised parton distribution functions. After doing so, we have a finite remainder which we can write in terms of $\rho_{ij}^{(sc)}(x)$ and $\rho^{(r)}_{ij}(x, y, \phi)$, following the notation of Eq.~\eqref{eq:reals}. The explicit expressions for each channel are given below as

\begin{align}
    \rho_{gg}^{(sc)} =& 2 \pi^2 \delta(1-x) +12 \left(1 - (1-x) x \right)^2 \left(2 \Duno - \frac{\log(x)}{1-x}\right),\\
    \rho_{qg}^{(sc)} = \rho_{gq}^{(sc)} =& \frac{4}{3} \left(x^2 + (2 \log(1-x) - \log(x)) (1 + (1-x)^2) \right),\\
    \rho^{(r)}_{ij} (x, y, \phi) =& \frac{1}{2 \pi s^2} \sum_{(kl)} \frac{{\rm Re}\left( (C_\text{LO}^{3H})^* C_\text{LO}^{2H}(\{k,l\}) \right)}{|C_\text{LO}^{3H}|^2} f_{ij}(\{k,l\}),
\end{align}
where $\sum_{(kl)}$ denotes a sum over the three distinct combinations of pair of Higgs bosons that we label $\{k,l\}$, and the $f(\{k,l\})$ functions are defined for each channel as
\begin{align}
    f_{gg}(\{k,l\}) =& \G(s,t_k,u_k,q_1,q_2,s_{kl},m_H^2)
                    + \G(s,t_k,u_k,\hat q_1,\hat q_2,m_H^2,s_{kl}) \\
    f_{qg}(\{k,l\}) =& \frac{-2}{9}(1-x) (1+y) \Q(s,t_k,u_k,q_1,q_2,s_{kl},m_H^2) \\
    f_{gq}(\{k,l\}) =& \frac{-2}{9}(1-x) (1-y)
    \Q(s,u_k,t_k,\hat q_2,\hat q_1,s_{kl},m_H^2)\\
    f_{qq}(\{k,l\}) =& \frac{-8}{27}(1-x)^2 (1-y^2)
    \Q(t_k,s,u_k,q_1,q_1-q_2+s_{kl}-u_k,s_{kl},m_H^2).
\end{align}
The invariants entering as arguments of $f$ are defined for each $\{k,l\}$ pair. For a given $\{k,l\}$, we call $p_1$ and $p_2$ the four-momenta of the incoming partons, $p_{hh} = p_k + p_l$ the sum of the four-momenta of the outgoing Higgs bosons labelled $k$ and $l$, $p_h$ the four-momenta of the other Higgs and $k$ the one of the emitted parton. Then we define the invariants as
\begin{equation}
\begin{aligned}
    s =& (p_1+p_2)^2 = x Q^2, & q_1 =& (p_1 - p_h)^2,\\
    t_k =& (p1-k)^2 = -\frac{s}{2} (1-x)(1-y), & q_2 =& (p_2 - p_{hh})^2,\\
    u_k =& (p2-k)^2 = -\frac{s}{2} (1-x)(1+y), & \hat q_1 =& (p_1 - p_{hh})^2 = m_H^2-q_1-s+s_{kl}-t_k,\\
    s_{kl} =& p_{kl}^2, & \hat q_2 =& (p_2 - p_{h})^2  = m_H^2-q_2-s+s_{kl}-u_k.
\end{aligned}
\end{equation}
The functions $\G$ and $\Q$ are regular in the limits $x\to1$ and $y\to \pm 1$. From the expressions in Equations (A9-A12) we can see explicitly that soft divergences  appear only in the $gg$ channel, as all others $f_{ij}$ vanish in this limit. The $gg$ channel also shows divergent behaviour in both collinear $y\to \pm 1$ limits.  The $qg$ and $gq$ channels only have singularities in $y\to 1$ and $y\to-1$ respectively, while the $qq$ channel is completely regular and free of any divergences. 

The analytic expressions for $\G$ and $\Q$ are
\begin{align}
    \G(s,t,u,q_1,q_2,m_1^2,m_2^2) =& \frac{g_1(s,t,u,q_1,q_2,m_1^2,m_2^2)}{s (m_2^2 - q_1 + q_2 - t)} + \frac{g_2(s,t,u,q_1,q_2,m_1^2,m_2^2)}{q_1 q_2} \nonumber\\+& \frac{g_3(s,t,u,q_1,q_2,m_1^2,m_2^2)}{q_2 t u (q_1 + s + t - m_1^2 - m_2^2)} ,\\
    \Q(s,t,u,q_1,q_2,m_1^2,m_2^2) =& \frac{-4}{t (m_2^2 - q_2 - s + m_1^2 - u)}\nonumber\\\times&
    \Bigg(m_1^2 s^2 t + m_2^2 s^2 t - s^3 t + m_1^4 t^2 - m_1^2 s t^2 + s^2 t^2 + q_2^2 (s^2 + s t + t^2)\nonumber\\& + m_1^2 m_2^2 t u - m_2^2 s t u - s^2 t u - 2 m_1^2 t^2 u + s t^2 u + m_2^4 u^2 + m_1^2 t u^2\nonumber\\& - s t u^2 + t^2 u^2 - t u^3 + q_1^2 (s + u)^2 - q_2 (2 s^2 t + s (m_1^2 t - t^2 - 2 m_2^2 u \nonumber\\& - t u) + t (2 m_1^2 t + u (m_2^2 - 2 t + u))) + q_1 (s^2 t + u (-(m_1^2 t) - 2 m_2^2 u \nonumber\\& + t u) + s (m_1^2 t - 2 m_2^2 u + 2 t u) - q_2 (2 s^2 - t u + s (t + 2 u)))\Bigg)\nonumber\\
    +&\frac{-4}{t q_2}\nonumber\\\times&
    \Bigg(m_1^4 t^2 + q_2^2 (s^2 + s t + t^2) + m_1^2 m_2^2 t u - m_1^2 t^2 u + m_2^4 u^2 - 2 m_2^2 t u^2 \nonumber\\&+ t^2 u^2 + q_1^2 (s + u)^2 + q_2 (s^2 t + s (-(m_1^2 t) + 2 m_2^2 u) + t (-2 m_1^2 t \nonumber\\&+ u (-m_2^2 + t + u))) - q_1 (u (m_1^2 t + 2 m_2^2 u - 2 t u) - s (m_1^2 t \nonumber\\&- 2 m_2^2 u + 2 t u) + q_2 (2 s^2 - t u + s (t + 2 u)))\Bigg),
\end{align}
with $g_1,\,g_2$ and $g_3$ being the following polynomials:
\begin{align}
    g_1(s,t,u,q_1,q_2,m_1^2,m_2^2) &= 4 m_1^2 q_2 s^3 t - 4 q_2^2 s^3 t + q_2 s^4 t - 4 m_1^4 s^2 t^2 + 8 m_1^2 q_2 s^2 t^2 - 4 q_2^2 s^2 t^2 \nonumber\\& + m_1^2 s^3 t^2 + 9 q_2 s^3 t^2 - 2 s^4 t^2 - 2 m_1^2 s^2 t^3 + 12 q_2 s^2 t^3 - 8 s^3 t^3 +  4 q_2 s t^4\nonumber\\& - 10 s^2 t^4 - 4 s t^5 + m_1^2 q_2 s^3 u - q_2^2 s^3 u + 4 q_2 s^4 u + m_1^4 s^2 t u - 2 m_1^2 q_2 s^2 t u \nonumber\\&- 5 q_2^2 s^2 t u + 2 m_1^2 s^3 t u + 14 q_2 s^3 t u - 8 s^4 t u +  4 m_1^2 q_2 s t^2 u - 4 q_2^2 s t^2 u \nonumber\\&+ 8 m_1^2 s^2 t^2 u + 25 q_2 s^2 t^2 u - 19 s^3 t^2 u - 4 m_1^4 t^3 u + 8 m_1^2 q_2 t^3 u - 4 q_2^2 t^3 u \nonumber\\&- m_1^2 s t^3 u + 11 q_2 s t^3 u - 24 s^2 t^3 u -  10 s t^4 u + 9 q_2 s^3 u^2 + 25 q_2 s^2 t u^2 \nonumber\\&- 12 s^3 t u^2 + 5 m_1^2 s t^2 u^2 + 11 q_2 s t^2 u^2 - 34 s^2 t^2 u^2 + 8 m_1^2 t^3 u^2 - 8 q_2 t^3 u^2 \nonumber\\&- 19 s t^3 u^2 + 10 q_2 s^2 u^3 +  10 q_2 s t u^3 - 11 s^2 t u^3 - 15 s t^2 u^3 - 4 t^3 u^3 \nonumber\\&+ 4 q_2 s u^4 - 4 s t u^4 + m_2^4 (s^2 (t - 4 u) u - 4 t u^3) - q_1^2 (4 s t u^2 + 4 t u^3 \nonumber\\&+ s^3 (t + 4 u) + s^2 u (5 t + 4 u)) +  m_2^2 (m_1^2 s^2 t^2 + 2 s^3 t^2 + 6 s^2 t^3 + 4 s t^4 \nonumber\\&- 8 m_1^2 s^2 t u + 7 s^3 t u + 17 s^2 t^2 u + 10 s t^3 u + m_1^2 s^2 u^2 + 4 s^3 u^2 + 28 s^2 t u^2 \nonumber\\&- 8 m_1^2 t^2 u^2 + 19 s t^2 u^2 + 7 s^2 u^3 + 15 s t u^3 + 8 t^2 u^3 + 4 s u^4 + q_1 (-2 s^2 (t \nonumber\\&- 4 u) u + 4 s t u^2 + 8 t u^3 + s^3 (t + 4 u)) - q_2 (4 s t u^2 - 8 t^2 u^2 + s^3 (t + 4 u)\nonumber\\& + s^2 (t^2 - 8 t u + u^2))) +  q_1 (8 t^2 (m_1^2 - u) u^2 - s^4 (t + 4 u) - s^3 (10 t^2 \nonumber\\&+ 23 t u + 11 u^2 + m_1^2 (4 t + u)) - s (4 t^4 + 10 t^3 u + 15 t u^3 + 4 u^4 + t^2 u (4 m_1^2 \nonumber\\&+ 23 u)) -    s^2 (10 t^3 + 30 t^2 u + 33 t u^2 + 11 u^3 + m_1^2 (t^2 - 8 t u + u^2)) \nonumber\\&+ q_2 (-8 t^2 u^2 + 5 s^3 (t + u) + 4 s t u (t + u) + s^2 (t^2 + 4 t u + u^2))),\\
    g_2(s,t,u,q_1,q_2,m_1^2,m_2^2) &= q_2 s (3 s - 2 t) t (-(m_1^2 t) + q_2 (s + t)) + (q_2^3 (s + t) (s + 2 t) \nonumber\\&+ q_2^2 (s^2 (-m_1^2 + s) + 3 s (-2 m_1^2 + s) t + 2 (-3 m_1^2 + s) t^2) - 2 m_1^2 t^2 (m_1^4 \nonumber\\&+ (s + t)^2) + q_2 t (2 s^3 - s^2 (m_1^2 - 9 t) + 2 t (3 m_1^4 + t^2) + s (3 m_1^4 - 2 m_1^2 t \nonumber\\&+ 4 t^2))) u - 2 m_2^6 t u^2 + t (2 t (-m_1^2 + s + t)^2 + q_2^2 (3 s + 2 t) + q_2 (5 s^2 \nonumber\\&+ 4 t (-m_1^2 + t) + s (-3 m_1^2 + 10 t))) u^2 + 2 t (q_2 (2 s + t) + t (-m_1^2 \nonumber\\&+ 2 s + 2 t)) u^3 + 2 t^2 u^4 + q_1^3 t (s + u) (s + 2 u) + q_1^2 (s t (s^2 + m_1^2 t \nonumber\\&- q_2 (2 s + t)) + (s^2 (-m_1^2 + q_2 + 3 s) + 3 s (-q_2 + s) t + (-2 m_1^2 + 2 q_2 \nonumber\\&+ 3 s) t^2) u + (s (-m_1^2 + q_2 + s) + 2 s t + 2 t^2) u^2 - 2 s u^3) + m_2^2 (-((q_1 \nonumber\\&- q_2) s t (q_1 s + m_1^2 t - q_2 (s + t))) - (-4 q_2 s^3 + s (6 q_1 (q_1 - q_2) \nonumber\\&+ (q_1 - 5 q_2) s) t + (q_2 s + q_1 (-4 m_1^2 + 4 q_2 + 3 s)) t^2) u - (q_2 (m_1^2 \nonumber\\&- q_2 - 7 s) s + q_1 s (-m_1^2 + q_2 + 3 s) + 2 (m_1^4 + 3 q_1^2 - 2 m_1^2 q_2 + q_2^2 + q_1 s\nonumber\\& - 3 q_2 s + s^2) t + 4 (q_1 + s) t^2 + 2 t^3) u^2 + 2 (q_1 s + q_2 s - 2 t (s\nonumber\\& + t)) u^3 - 2 t u^4) + q_1 (s t (q_2^2 (s + t) + m_1^2 (4 s^2 + 7 s t + 2 t^2)\nonumber\\& - q_2 (8 s^2 + 5 s t + t (m_1^2 + 2 t))) + s (-2 q_2 s (-m_1^2 + q_2 + 4 s) + (-3 m_1^4\nonumber\\& + 6 m_1^2 q_2 - 3 q_2^2 + 5 m_1^2 s - 14 q_2 s + 2 s^2) t + (6 m_1^2 - 5 q_2 + 5 s) t^2 + 4 t^3) u \nonumber\\& + (q_2 (m_1^2 - q_2 - 5 s) s + (2 m_1^4 + 2 q_2^2 - 5 q_2 s + 9 s^2 - m_1^2 (4 q_2 + s)) t \nonumber\\&+ 10 s t^2 + 2 t^3) u^2 + 2 (-(q_2 s) + 2 t (s + t)) u^3 + 2 t u^4) + m_2^4 t u (-3 q_2 s \nonumber\\&- 2 m_1^2 t + 2 q_2 t + 2 t u + 3 q_1 (s + 2 u)),\\
    g_3(s,t,u,q_1,q_2,m_1^2,m_2^2) &= -(q_1^3 u (s + u)^2 (3 t^2 + 4 s u)) + q_2^3 t (s + t)^2 (4 s t + 3 u^2) + q_2^2 (-4 s^4 u^2 \nonumber\\&+ s^3 u (-5 t^2 + 4 (m_1^2 + m_2^2) u - 10 t u) + t^2 u^2 (m_2^2 (t - 3 u) + t (-6 m_1^2\nonumber\\& + t + 5 u)) + s^2 t (u (8 m_2^2 t + 4 m_2^2 u - 11 t u + u^2) + m_1^2 (-8 t^2\nonumber\\& + 4 t u + 5 u^2)) + s t (-(m_1^2 t (8 t^2 - 4 t u + u^2)) + u (5 t^3 - 4 t^2 u + 3 m_2^2 u^2 \nonumber\\&+ t u (5 m_2^2 + 6 u)))) + q_1^2 (q_2 (-3 t^3 u^2 + s t^2 u (3 t + 11 u) + 4 s^3 (t^2 + 2 u^2)\nonumber\\& + s^2 u (11 t^2 + 8 t u + 8 u^2)) + u (-4 s^4 u - s^3 (4 t^2 + 9 t u - 4 (m_1^2\nonumber\\& - 2 u) u) + t^2 u (u (-7 t + u) + 3 m_1^2 (t + u)) - s^2 (5 t^3 + 7 t^2 u + 16 t u^2 \nonumber\\&+ 4 u^3 + m_1^2 (-3 t^2 + 4 t u - 8 u^2)) + 2 m_2^2 (s + u) (2 s^2 u + 5 t^2 u + s (t^2 \nonumber\\&+ 6 u^2)) + s (-(t u (12 t^2 + 2 t u + 7 u^2)) + m_1^2 (-3 t^3 + 6 t^2 u + 4 t u^2 \nonumber\\&+ 4 u^3)))) + u (m_2^6 u (-(s t^2) + 4 t^2 u + 4 s u^2) + m_2^2 t (s^3 (4 m_1^2 t + u (t \nonumber\\&+ 5 u)) + s^2 (m_1^2 (6 t^2 + 4 t u - 9 u^2) + u (5 t^2 + 10 t u + 3 u^2)) \nonumber\\&+ t u (4 m_1^4 t + 2 t (3 t - u) u + m_1^2 (-3 t^2 + 2 t u + 5 u^2)) + s (t u (7 t^2 \nonumber\\&+ 5 t u + u^2) + m_1^4 (t^2 + 6 t u + 4 u^2) + m_1^2 (3 t^3 + 5 t^2 u - 3 t u^2\nonumber\\& - 3 u^3))) + m_2^4 (-(s^2 u^2 (5 t + 4 u)) + t^2 u (u (-9 t + u) + 4 m_1^2 (t \nonumber\\&+ u)) + s (-(t u (2 t^2 + 4 t u + 7 u^2)) + m_1^2 (t^3 - t^2 u + 4 t u^2 + 4 u^3)))\nonumber\\& - t^2 (4 m_1^2 s^4 + s^3 (-4 m_1^4 + u (t + 5 u) + m_1^2 (11 t + 5 u)) + s^2 (m_1^4 (-7 t\nonumber\\& + u) + m_1^2 (11 t^2 + 16 t u - 4 u^2) + u (5 t^2 + 5 t u - u^2)) + (-m_1^2 \nonumber\\&+ t) u (t (t - u) u + m_1^2 (t^2 + 7 t u + 4 u^2)) + s (-6 m_1^6 u + t u (4 t^2 + t u \nonumber\\&- 2 u^2) - m_1^4 (4 t^2 + 6 t u + u^2) + m_1^2 (4 t^3 + 9 t^2 u + 8 t u^2 + 5 u^3)))) \nonumber\\&+ q_2 (8 s^5 t u + s^4 t u (-8 m_1^2 - 8 m_2^2 + 29 t + 25 u) - s^3 u (t (21 m_1^2 t\nonumber\\& - 52 t^2 + 17 m_1^2 u - 76 t u - 23 u^2) + m_2^2 (20 t^2 + 25 t u + 8 u^2)) \nonumber\\&+ t u (3 m_1^4 t^2 u - m_2^4 t u (4 t + u) - 2 m_1^2 (2 t^4 + 4 t^3 u + 8 t^2 u^2 \nonumber\\&+ 5 t u^3 + 2 u^4) + t (4 t^4 + 11 t^3 u + 18 t^2 u^2 + 14 t u^3 + 4 u^4) - m_2^2 (4 t^4 \nonumber\\&+ 7 t^3 u + 4 u^4 + t u^2 (-3 m_1^2 + 14 u) + t^2 u (5 m_1^2 + 17 u))) - s^2 u (m_2^4 (t^2\nonumber\\& - 8 u^2) + t (-45 t^3 + 4 m_1^4 u - 103 t^2 u - 63 t u^2 - 12 u^3 + m_1^2 (26 t^2 \nonumber\\&+ 27 t u + 17 u^2)) + m_2^2 (m_1^2 (t^2 + 4 t u - 8 u^2) + t (29 t^2 + 66 t u + 29 u^2))) \nonumber\\&+ s t (m_1^4 t (4 t^2 - 4 t u - 13 u^2) - m_1^2 u (19 t^3 + 22 t^2 u + 28 t u^2 \nonumber\\&+ 10 u^3 + m_2^2 (t^2 + 4 t u + 3 u^2)) + u (18 t^4 + 57 t^3 u + 58 t^2 u^2 + 29 t u^3 \nonumber\\&+ 4 u^4 - m_2^4 (t^2 - 14 t u + 4 u^2) - m_2^2 (13 t^3 + 56 t^2 u + 35 t u^2 \nonumber\\&+ 11 u^3)))) - q_1 (q_2^2 (-3 t^2 u^3 + s t u^2 (11 t + 3 u) + 4 s^3 (2 t^2 + u^2) \nonumber\\&+ s^2 t (8 t^2 + 8 t u + 11 u^2)) - q_2 (8 s^4 u (t + u) + s^3 u (29 t^2 + 39 t u\nonumber\\& + 8 u (-m_1^2 - m_2^2 + u)) + t u (4 t^4 + 8 t^3 u + 4 u^4 + t u^2 (-3 m_1^2 \nonumber\\&+ m_2^2 + 14 u) + t^2 u (3 m_1^2 + 7 m_2^2 + 18 u)) + s t u (-3 m_1^2 t^2 + 17 t^3 \nonumber\\&+ 14 m_1^2 t u + 53 t^2 u + 3 m_1^2 u^2 + 37 t u^2 + 11 u^3 - 2 m_2^2 (t^2 + 13 t u \nonumber\\&- 2 u^2)) + s^2 (4 t u (10 t^2 + 17 t u + 7 u^2) - 2 m_2^2 u (5 t^2 + 2 t u + 8 u^2)\nonumber\\& + m_1^2 (8 t^3 - 7 t^2 u - t u^2 - 8 u^3))) + u (m_2^4 (11 t^2 u^2 - s^2 (t^2 - 8 u^2) \nonumber\\&+ s (5 t^2 u + 12 u^3)) - m_2^2 (s^3 u (5 t + 8 u) + t^2 u (2 (8 t - u) u - 7 m_1^2 (t + u)) \nonumber\\&+ s^2 (t^3 + 8 t^2 u + 21 t u^2 + 8 u^3 + m_1^2 (t^2 + 4 t u - 8 u^2)) + s (t u (15 t^2 \nonumber\\&+ 6 t u + 14 u^2) + 2 m_1^2 (t^3 - 3 t^2 u - 4 t u^2 - 4 u^3))) + t (s^4 (t \nonumber\\&+ 5 u) + s^3 (3 m_1^2 t + 3 t^2 - m_1^2 u + 6 t u + 8 u^2) + s^2 (2 t^3 - 4 m_1^4 u \nonumber\\&+ 8 t^2 u + 12 t u^2 + 3 u^3 + 4 m_1^2 (2 t^2 + 2 t u - 3 u^2)) + t u (3 m_1^4 t \nonumber\\&+ t (5 t - 2 u) u + m_1^2 (-2 t^2 + 3 t u + 5 u^2)) + s (t u (10 t^2 + 3 t u + u^2) \nonumber\\&+ m_1^4 (-3 t^2 + 6 t u + 4 u^2) + m_1^2 (7 t^3 + 3 t^2 u - 2 t u^2 - 3 u^3))))).
\end{align}
\end{appendices}
\end{document}